\documentclass[journal]{IEEEtran}

\usepackage{lineno,hyperref}

\usepackage{mathtools}
\usepackage{amsmath}
\usepackage{array}
\usepackage{mathrsfs}
\usepackage{amssymb}
\usepackage{booktabs}
\usepackage{algorithm}
\usepackage{algpseudocode}
\usepackage{multirow}
\usepackage{tabularx}
\usepackage{tabulary}
\usepackage{dcolumn}
\usepackage{threeparttable}
\usepackage{array}
\usepackage{bbm}
\usepackage{color, soul}

\modulolinenumbers[5]

\usepackage{scalerel}
\usepackage{tikz}
\usetikzlibrary{svg.path}

\definecolor{orcidlogocol}{HTML}{A6CE39}
\tikzset{
	orcidlogo/.pic={
		\fill[orcidlogocol] svg{M256,128c0,70.7-57.3,128-128,128C57.3,256,0,198.7,0,128C0,57.3,57.3,0,128,0C198.7,0,256,57.3,256,128z};
		\fill[white] svg{M86.3,186.2H70.9V79.1h15.4v48.4V186.2z}
		svg{M108.9,79.1h41.6c39.6,0,57,28.3,57,53.6c0,27.5-21.5,53.6-56.8,53.6h-41.8V79.1z M124.3,172.4h24.5c34.9,0,42.9-26.5,42.9-39.7c0-21.5-13.7-39.7-43.7-39.7h-23.7V172.4z}
		svg{M88.7,56.8c0,5.5-4.5,10.1-10.1,10.1c-5.6,0-10.1-4.6-10.1-10.1c0-5.6,4.5-10.1,10.1-10.1C84.2,46.7,88.7,51.3,88.7,56.8z};
	}
}

\newcommand\orcidicon[1]{\href{https://orcid.org/#1}{\mbox{\scalerel*{
				\begin{tikzpicture}[yscale=-1,transform shape]
				\pic{orcidlogo};
				\end{tikzpicture}
			}{|}}}}

\usepackage{hyperref} 

\begin{document}


\title{A Novel Smoothed Loss and Penalty Function for Noncrossing Composite Quantile Estimation via Deep Neural Networks}

\author{
	Kostas~Hatalis~\orcidicon{0000-0001-8191-5728},~\IEEEmembership{Student Member,~IEEE,}
	Alberto~J.~Lamadrid~\orcidicon{0000-0003-2539-9487},~\IEEEmembership{Senior Member,~IEEE,}
	Katya~Scheinberg,
	and~Shalinee~Kishore,~\IEEEmembership{Senior Member,~IEEE}
	\thanks{K. Hatalis and S. Kishore are with the Department
		of Electrical and Computer Engineering, Lehigh University, Bethlehem, PA, 18015, USA (e-mails: {\tt \{kmh511,shk2\}@lehigh.edu}).}
	\thanks{K. Scheinberg is with the Department of Industrial Engineering and System Engineering, Lehigh University, Bethlehem, PA, 18015, USA	(e-mail: {\tt katyas@lehigh.edu}).}
	\thanks{A. J. Lamadrid is with the Department of Economics and the Department of Industrial and System Engineering, Lehigh University, Bethlehem, PA, 18015 USA (e-mail: {\tt ajlamadrid@ieee.org})}
}


\maketitle

\begin{abstract}
	Uncertainty analysis in the form of probabilistic forecasting can significantly improve decision making processes in the smart power grid when integrating renewable energy sources such as wind. Whereas point forecasting provides a single expected value, probabilistic forecasts provide more information in the form of quantiles, prediction intervals, or full predictive densities. Traditionally quantile regression is applied for such forecasting and recently quantile regression neural networks have become popular for weather and renewable energy forecasting. However, one major shortcoming of composite quantile estimation in neural networks is the quantile crossover problem. This paper analyzes the effectiveness of a novel smoothed loss and penalty function for neural network architectures to prevent the quantile crossover problem. It's efficacy is examined on the wind power forecasting problem. A numerical case study is conducted using publicly available wind data from the Global Energy Forecasting Competition 2014. Multiple quantiles are estimated to form 10\%, to 90\% prediction intervals which are evaluated using a quantile score and reliability measures. Benchmark models such as the persistence and climatology distributions, multiple quantile regression, and support vector quantile regression are used for comparison where results demonstrate the proposed approach leads to improved performance while preventing the problem of overlapping quantile estimates.
\end{abstract}

\begin{IEEEkeywords}
	Nonparametric probabilistic forecasting, prediction intervals, multiple quantile regression, deep neural networks.
\end{IEEEkeywords}

\IEEEpeerreviewmaketitle

\section{Introduction}

\IEEEPARstart{I}{n} the last thirty years wind power has experienced rapid global growth, and in some countries, it is the most used form of renewable energy. However, due to the chaotic nature of the weather, variable and uncertain wind power production poses planning and operational challenges unseen in conventional generation. From the grid operator's perspective, uncertainty in wind production could cause inefficiencies in the power flow, operating reserve requirements, stochastic unit commitment, and electricity market settlements \cite{makarov2009operational, bukhsh2016integrated, botterud2012wind}. From the wind generator's perspective, reliable wind forecasts are needed for several operations at a wind farm, ranging from energy storage control to bidding and trading in energy markets. Thus, to ensure both stable grid operations and continued growth and increased penetration of wind power, highly reliable forecasting of wind power production is needed.

Traditionally wind power prediction has focused on developing point forecasts which provide a single expected output for a given look-ahead time. Point forecasting horizons fall into several scales: very short-term (seconds or minutes ahead), short-term (hours to days ahead), long-term (weeks or months ahead), and seasonal. A thorough review in wind forecasting can be found in \cite{monteiro2009wind}. However, point forecasting can result in certain errors which can be significant and they also lack information on uncertainty. Therefore, a significant research effort has begun recently by the renewables forecasting community \cite{hong2016probabilistic} to produce fully probabilistic predictions which derive quantitative information on the associated uncertainty of power output. For example, to capture the uncertainty of wind power, forecasting errors can be statistically analyzed and modeled by the Beta distribution. However, such assumption may not be applicable for short-term forecasting, and thus researchers are looking at different approaches for probabilistic wind power forecasts by quantifying prediction uncertainty. Although there are various methods proposed, it is still a challenge to make accurate and reliable probabilistic predictions for volatile renewables, such as wind. 

Probabilistic forecast can play a key role in integrating and managing wind farms. For instance, in \cite{doherty2005new} the optimal level of generation reserves is estimated using the uncertainty of wind power predictions, and in \cite{usaola2004benefits,castronuovo2004optimization} the optimization of wind energy production is investigated taking into account the forecasts of a probabilistic prediction method. Additionally, increased revenues can be obtained using bidding strategies built on predictive densities, as shown in \cite{bathurst2002trading,pinson2007trading}. Wind power density forecasting can be used for analysis of probabilistic load flow, as in \cite{karakatsanis1994probabilistic}. 


Our work is motivated by exploring a direct and nonparametric probabilistic forecasting approach for wind power. To address the problem of dealing with nonlinearity in wind data \cite{lange2005uncertainty}, we propose a novel neural network model which we call the smooth pinball neural network (SPNN). This network is able to provide probabilistic forecasts in the form of multiple monotonically increasing quantiles estimated simultaneously. The main contributions of our approach can be summarized as follows:

\begin{enumerate}
	\item We propose and investigate a new objective function which is a logistic based smooth approximation of the pinball loss function for multiple quantile regression.
	\item We introduce a smooth penalty scheme to prevent the quantile crossover problem.
	\item We showcase how a multiple quantile based neural network can be used for probabilistic forecasting of wind.
	\item We design experiments to validate our model using publicly available data from 10 wind farms from the Global Energy Forecasting Competition 2014 and benchmark performance with common and advanced methods.
	\item We show our method improves the skill, reliability, and sharpness of forecasts over various benchmarks.
\end{enumerate}

In Section \ref{s:related}, we provide a literature review of nonparametric probabilistic forecasting approaches for wind power and we dive deeper into related work on QR, quantile based neural networks, and approaches for preventing what is known as the quantile cross over problem. In Section \ref{s:prob}, we provide the mathematical background on probabilistic forecasting, QR, and evaluation methods. Section \ref{s:mod} overviews our model, its architecture, and training. Results and discussion of our case study are presented in Section \ref{s:res}. We conclude the paper and review future research directions in Section \ref{s:con}.

\section{Related Work} \label{s:related}

\subsection{Probabilistic Forecasting of Wind Power}

Over the last several years there has been a large body of work conducted in nonparametric probabilistic forecasting of wind power as well as other renewables such as solar power. Recently the Global Energy Forecasting Competition in 2014 \cite{hong2016probabilistic} and 2017 are further proof of the rising interest in probabilistic forecasting. Probabilistic wind models are either meteorological ensembles that are obtained by a weather model \cite{giebel2003using} or are statistical methods \cite{foley2012current}. Under the statistical approach, we can estimate full predictive distributions in the form of quantiles or prediction intervals (PIs). One PI estimation scheme is shown in \cite{sideratos2012probabilistic} which uses a radial basis function neural network.  

Some of the most recent forecasting methods include extreme learning machines \cite{wan2017direct} where a direct quantile regression approach was presented to efficiently generate nonparametric probabilistic forecasting of wind power generation combining extreme learning machine and quantile regression. Hybrid intelligent methods have also been explored in \cite{haque2014hybrid} by feeding deterministic wind power forecasts made by a combination of wavelet transform and fuzzy ARTMAP network, optimized by using firefly optimization algorithm, in quantile regression. Another approach to forecast the density of wind power is to take an ensemble of point forecasts and calculate the mean and variance of the combined forecasts. This has been studied in \cite{wang2017deep} where a wavelet transform and a convolutional neural network are used for ensemble point forecasting. Another ensemble approach can be seen in \cite{taylor2009wind} where time series models such as ARMA and GARCH are combined to form density forecasts. One of the most prevalent approaches to probabilistic forecasting of wind power is to apply quantile regression (QR) which can be used to estimate different wind power quantiles \cite{zhang2014review}. 

Another alternative to nonparametric probabilistic wind forecasting is the application of the Lower Upper Bound Estimation (LUBE) method \cite{khosravi2011lower}. The LUBE method constructs a neural network with two outputs for estimating the prediction interval bounds. The coverage width-based criterion is used as the loss function for estimating PIs, and simulated annealing or particle swarm optimization \cite{quan2014short} can be used to minimize that loss function. A complete review on probabilistic forecasting of wind power can be found in \cite{zhang2014review}. Other reviews on probabilistic forecasting methods can be found in \cite{van2018review} for solar power, \cite{hong2016probabilisticload} for load forecasting, and \cite{nowotarski2017recent} for electricity price forecasting.

\subsection{Nonlinear Quantile Regression}

Here we provide a thorough review of QR methods, particularly nonlinear versions, as background to our proposed method. There are many variations of QR which are traditionally solved using linear programming algorithms. In \cite{bremnes2004probabilistic} local QR is applied to estimate different quantiles, while in \cite{nielsen2006using} a spline-based QR is used to estimate quantiles of wind power. In \cite{landry2016probabilistic} quantile loss gradient boosted machines are used to estimate many quantiles and in \cite{juban2016multiple} multiple quantile regression is used to predict a full distribution with optimization achieved by using the alternating direction method of multipliers. Quantile regression forests \cite{juban2008uncertainty} are another approach in forecasting which are an extension of regression forests based on classification and regression trees. 

With QR being a comprehensive strategy for providing the conditional distribution of a response $y$ given $x$, we highlight several of its variants. In a generalization of QR \cite{powell1986censored} introduce the censored QR model, which consistently estimates conditional quantiles when observations on the dependent variable are censored. Yu and Jones \cite{yu1998local} propose a nonparametric version of QR estimation by using a kernel-weighted  local linear fitting. Chen et al. \cite{chen2009copula} propose a copula-based nonlinear quantile autoregression, addressing the possibility of deriving nonlinear parametric models for different conditional quantile functions. QR can also be hybridized with machine learning methods to form powerful nonlinear models. For instance, support vector regression is introduced for QR in  \cite{hwang2005simple}, yielding support vector quantile regression (SVQR). SVQR extends the QR model to non-linear and high dimensional spaces, but it requires solving a quadratic programming problem. 

Due to their flexibility in modeling elaborate nonlinear data sets, artificial neural networks are another dominant class of machine learning algorithms which are also very popular for renewable forecasting \cite{hatalis2014multi,azad2014long}. Taylor \cite{taylor2000quantile} is the first to propose a quantile regression neural network (QRNN) method, combining the advantages of both QR and a neural network. This method can reveal the conditional distribution of the response variable and can also model the nonlinearity of different systems. The author applies this method to estimate the conditional distribution of multi-period returns in financial systems, which avoids the need to specify the explanatory variables explicitly. However, the paper does not address how the network was optimized. The same QRNN was later used by \cite{feng2010robust} for credit portfolio data analysis where results showed that QRNN is more robust in fitting outliers compared to both local linear regression and spline regression. In \cite{xu2016quantile} an autoregressive version of QRNN is used for applications to evaluating value at risk, and \cite{cannon2011quantile} implements the QRNN model in R as a statistical package. 

In all the QR approaches mentioned, only a single quantile is estimated at a time. In the case of estimating multiple quantiles, this could lead to what is known as the quantile crossover problem, where a lower quantile overlaps a higher quantile.  Equivalently, a prediction interval for a lower probability (e.g., range in which 10\% of future values are predicted to lie) exceeds that of a higher probability (e.g., the range in which 20\% of the future values are predicted to lie). Crossing quantiles are undesirable as it violates the principle of cumulative distribution functions where their associated inverse functions should be monotonically increasing. A possible way to prevent this issue is to utilize simple heuristics of reordering estimated quantiles. However, this approach does not have a strong theoretical foundation and may lead to inappropriate quantiles. The solution then is to optimize quantiles together with non-crossing constraints. In \cite{takeuchi2006nonparametric,hatalis2017empirical} a constrained support vector quantile regression (CSVQR) method is developed with non-crossing constraints where it was used to fit quantiles on static data. However, CSVQR is computationally very expensive and slow to train. In Section \ref{s:mod} we review approaches for preventing the quantile crossover problem in neural networks and we also propose a novel way to prevent this problem using a smooth penalty function.

\section{Probabilistic Forecasting} \label{s:prob}

This section highlights the underlying mathematics in probabilistic forecasting, overviews linear quantile regression, and summarizes the main evaluation metrics for density forecasts. Given a random variable $Y_t$ such as wind power at time $ t $, its probability density function is defined as $f_t$ and its the cumulative distribution function as $F_t$. If $ F_{t} $ is strictly increasing, the quantile $ q_{t}^{(\tau)} $ of the random variable $ Y_{t} $ with nominal proportion $ \tau $ is uniquely defined on the value $ x $ such that $ P(Y_t < x) = \tau $. It can also be defined as the inverse of the distribution function $ q_{t}^{(\tau)} = F_{t}^{-1}(\tau) $. A quantile forecast  $ \hat{q}_{t+z}^{(\tau)} $ is an estimate of the true quantile $  q_{t+z}^{(\tau)} $ for the lead time $ t+z $, given a predictor values (such as numerical wind speed forecasts). Prediction intervals are another type of probabilistic forecast and give a range of possible values within which an observed value is expected to lie with a certain probability $ \beta \in [0,1] $. A prediction interval $ \hat{I}^{(\beta)}_{t+z} $ is defined by its lower and upper bounds, which are the quantile forecasts $ \hat{I}^{(\beta)}_{t+z} = \left[ \hat{q}_{t+z}^{(\tau_{l})} ,\hat{q}_{t+z}^{(\tau_{u})} \right] = \left[ l_{t}^{(\beta)},u_{t}^{(\beta)} \right] $ whose nominal proportions $ \tau_l $ and $ \tau_u $ are such that $ \tau_u - \tau_l = 1-\beta $. 

In probabilistic forecasting, we are trying to predict one of two classes of density functions, either parametric or nonparametric. When the future density function is assumed to take a certain distribution, such as the Normal distribution, then this is called parametric probabilistic forecasting. For a nonlinear and bounded process such as wind generation, probability distributions of future wind power, for instance, may be skewed and heavy-tailed distributed \cite{dorvlo2002estimating}. Else if no assumption is made about the shape of the distribution, a nonparametric probabilistic forecast $ \hat{f}_{t+z} $ \cite{pinson2007non} can be made of the density function by gathering a set of $ M $ quantiles forecasts such that $ \hat{f}_{t+z} = \left\lbrace  \hat{q}_{t+z}^{(\tau_{m})} ,m=1,...,M|0\leq \tau_1 < ... < \tau_M \leq 1 \right\rbrace $ with chosen nominal proportions spread on the unit interval. In this paper, we consider forecasting wind power on the resolution of one hour (predicting outwards to a month worth of values). On this resolution scale of an hour, the wind density may fluctuate therefore making nonparametric forecasting more ideal then fitting a parametric density \cite{zhang2014review}.

\subsection{Quantile regression}

Quantile regression is a popular approach for nonparametric probabilistic forecasting. Koenker and Bassett \cite{koenker1978regression} introduce it for estimating conditional quantiles and is closely related to models for the conditional median \cite{koenker2005quantile}. Minimizing the mean absolute function leads to an estimate of the conditional median of a prediction. By applying asymmetric weights to errors through a tilted transformation of the absolute value function, we can compute the conditional quantiles of a predictive distribution. The selected transformation function is the pinball loss function as defined by
\begin{equation} \label{pinball}
\rho_{\tau}(u) = \left\lbrace 
\begin{array}{cl}
\tau u    & \mbox{if } u \geq 0 \\
(\tau-1)u & \mbox{if } u < 0
\end{array} \right.,
\end{equation}

\noindent where $ 0 < \tau < 1 $ is the tilting parameter. To better understand the pinball loss, we look at an example for estimating a single quantile. If an estimate falls above a reported quantile, such as the 0.05-quantile, the loss is its distance from the estimate multiplied by its probability of 0.05. Otherwise, the loss is its distance from the realization multiplied by one minus its probability (0.95 in the case of the 0.05-quantile). The pinball loss function penalizes low-probability quantiles more for overestimation than for underestimation and vice versa in the case of high-probability quantiles. Given a vector of predictors $ X_t $ where $ t = 1,...,N $, a vector of weights $ W $ and intercept $ b $ coefficient in a linear regression fashion, the conditional $ \tau $ quantile $ \hat{q}_\tau $ is given by $ \hat{q}_{t}^{(\tau)} = W^\top X_t +b $. To determine estimates for the weights and intercept we solve the following minimization problem
\begin{equation} \label{eq3}
\min_{W,b} \frac{1}{N} \sum_{t=1}^{N} \rho_\tau (y_t-\hat{q}_{t}^{(\tau)}),
\end{equation}

\noindent where $ y_t $ is the observed value of the predictand. The formulation above in Eq. (\ref{eq3}) can be minimized by a linear program. 

\subsection{Evaluation Metrics}

In probabilistic forecasting it is essential to evaluate the quantile estimates and if desired also evaluate derived predictive intervals. Therefore, we use as evaluation measures the quantile score, interval reliability, and interval sharpness. To evaluate quantile estimates, one can use the pinball function directly as an assessment called the quantile score (QS). We choose QS as our main  evaluation measure for the following reasons. When averaged across many quantiles it can evaluate full predictive densities; it is found to be a proper scoring rule \cite{grushka2017quantile}; it is related to the continuous rank probability score; and it is also the main evaluation criteria in the 2014 Global Energy Forecasting Competition (GEFCOM 2014), the source of our testing data. QS calculated overall $ N $ test observations and $ M $ quantiles is defined as
\begin{equation*}
QS = \sum_{t=1}^{N} \sum_{m=1}^{M} \rho_{\tau_m} (y_t - \hat{q}_{t}^{(\tau_{m})})
\end{equation*}

\noindent where $ y_t $ is an observation used to forecast evaluation such future wind power observations. To evaluate full predictive densities, QS is averaged across all target quantiles for all look ahead time steps using equal weights. A lower QS indicates a better forecast. 

With the QS calculated we can then also see what the relative performance of SPNN is with respect to some benchmark method. We can assess relative performance between methods using the quantile verification skill score (QVSS) \cite{friederichs2007statistical}
\begin{equation*}
QVSS = 1-\frac{QS^{for}}{QS^{ref}}
\end{equation*}
\noindent where $QS^{for}$ is QS for the forecast method of interest (SPNN in our case), and $ QS^{ref} $ is the QS value for the reference forecast of a benchmark method, which we will assume to be linear quantile regression. If QVSS is positive then forecast of interest performs better than the reference forecast, and a QVSS = 1 means a perfect forecast. Negative QVSS values indicate that forecast of interest performs worse than the reference forecast.

In some applications, it may be needed to have wind forecasts in the form of prediction intervals (PIs) and as such, we look at two secondary evaluation measures: reliability and sharpness. Reliability is a measure which states that over an evaluation set the observed and nominal probabilities should be as close as possible, and the empirical coverage should ideally equal the preassigned probability. Sharpness is a measure of the width of prediction intervals, defined as the difference between the upper $ u_{t}^{\beta_i} $ and lower $ l_{t}^{\beta_i} $ interval values. For interval reliability we use the average coverage error (ACE) metric \cite{zhang2014review} and for measuring interval sharpness we use the interval score (IS) which can also be used to evaluate the overall skill of PIs \cite{gneiting2007strictly}. For measuring reliability, PIs show where future wind power observations are expected to lie, with an assigned probability termed as the PI nominal confidence (PINC) $ 100(1-\beta_i)\% $. Here $ i = 1...M/2 $ indicates a specific coverage level. The coverage probability of estimated PIs is expected to eventually reach a nominal level of confidence over the test data. A measure of reliability which shows target coverage of the PIs is the PI coverage probability (PICP), which is defined by
\begin{equation*}
PICP_i=\frac{1}{N}\sum_{t=1}^{N} \mathbbm{1}\{y_t \in I_{t}^{\beta_i}({x}_t)\}.
\end{equation*}

For reliable PIs, the examined PICP should be close to its corresponding PINC. A related and easier to visualize assessment index is the average coverage error (ACE), which is defined by
\begin{equation*}
ACE = \sum_{\ i = 1}^{M/2} |PICP_i-100(1-\beta_i)|.
\end{equation*}

\noindent This assumes calculation across all test data and coverage levels. To ensure PIs have high reliability, the ACE should be as close to zero as possible. A high reliability can be easily achieved by increasing or decreasing the distance between lower and upper interval bounds. Thus, the width of a PI can also influence its quality. For measuring the effective width of PIs we use the sharpness score proposed by \cite{pinson2007non} which measures how wide PIs are by focusing on the mean size of the intervals only. We define $\hat{q}^{u}_{t}-\hat{q}^{l}_{t}$ as the size of the central interval forecast with nominal coverage rate $ (1 - \beta) $. For lead times $t = 1...N_{test}$, a measure of sharpness for PIs is then given by the mean size of the intervals
\begin{equation*} \label{e:sharp}
Sharpness=\frac{1}{N_{test}}\sum_{t=1}^{N_{test}}(\hat{q}^{u}_{t}-\hat{q}^{l}_{t}).
\end{equation*}

\noindent A lower sharpness score is considered more ideal, but too small and the PIs would not cover enough of the observed data. Thus sharpness is typically a measure to be considered along with reliability and a skill score. QS is a score that measures the skill of individual quantiles; to measure the skill of individual PIs we apply the interval score (IS) \cite{gneiting2007strictly}. The IS - when evaluated with all test data and coverage levels - is defined by
\begin{equation*} 
\begin{multlined}
IS = \frac{2}{NM} \sum_{t = 1}^{N} \sum_{\ i= 1}^{M/2}
(u_{t}^{\beta_i} - l_{t}^{\beta_i}) +\\ \frac{2}{\beta_i}(l_{t}^{\beta_i}-y_t) \mathbbm{1}\{y_t < l_{t}^{\beta_i}\} +  \frac{2}{\beta_i}(y_t - u_{t}^{\beta_i}) \mathbbm{1}\{y_t > u_{t}^{\beta_i}\}
\end{multlined}.
\end{equation*}

\noindent The prediction model is rewarded for narrow PIs and is penalized if the observation misses the interval. The size of the penalty depends on $ \beta_i $. Including all aspects of PI evaluation, the IS can be used to compare the overall skill and sharpness of interval forecasts. However, IS cannot identify the contributions of reliability and sharpness to the overall skill. Thus, ACE and sharpness are both used for evaluation of PIs along with QS for evaluation of quantile estimation.

\section{Smooth Pinball Network Model}\label{s:mod}

We propose to use a feedforward neural networks for probabilistic forecasting due to their flexibility and strength in dealing with nonlinear and nonstationary data. We can use the pinball loss in the objective function of such a neural network to estimate conditional quantiles. However, the pinball function $ \rho $ employed by the original linear quantile regression model in Eq. \eqref{pinball} is not differentiable at the origin, $ x = 0 $. The non-differentiability of $ \rho $ makes it difficult to apply gradient-based optimization methods in fitting the quantile regression model. Gradient-based methods are preferable for training neural networks since they are time efficient, easy to implement and yield a local optimum. Therefore, we need a smooth approximation of the pinball function that allows for the direct application of gradient-based optimization. We call our new model the smooth pinball neural network (SPNN). 

We are not the first to apply a smooth approximation to the pinball function for a quantile regression based neural network.  \cite{cannon2011quantile} used the Huber norm to construct smooth approximations of the pinball loss function, following the work in \cite{chen2007finite}, to form a QRNN. Using the same Huber norm approximation, a composite QRNN is proposed in \cite{xu2017composite} to estimate multiple quantiles. The Huber norm requires multiple optimization runs with a fixed schedule of a decreasing smoothing constant to from the final weights and biases. Chen et al. \cite{chen1996class} introduced another class of smooth functions for nonlinear optimization problems and applied this idea to support vector machines \cite{lee2001ssvm}. Emulating the work of Chen, a study by Zheng \cite{zheng2011gradient} presents an approximation to the pinball loss function by a smooth logistic function; this then allows the application of gradient descent for optimization. Zheng called the resulting algorithm the gradient descent smooth quantile regression model. We extend that model here for the case of a neural network. Based on our knowledge, we are the first to investigate the usage of a smooth logistic loss function to estimate multiple quantile using a neural network.

\begin{figure}[t]
	\centering
	\includegraphics[width=.5 \textwidth]{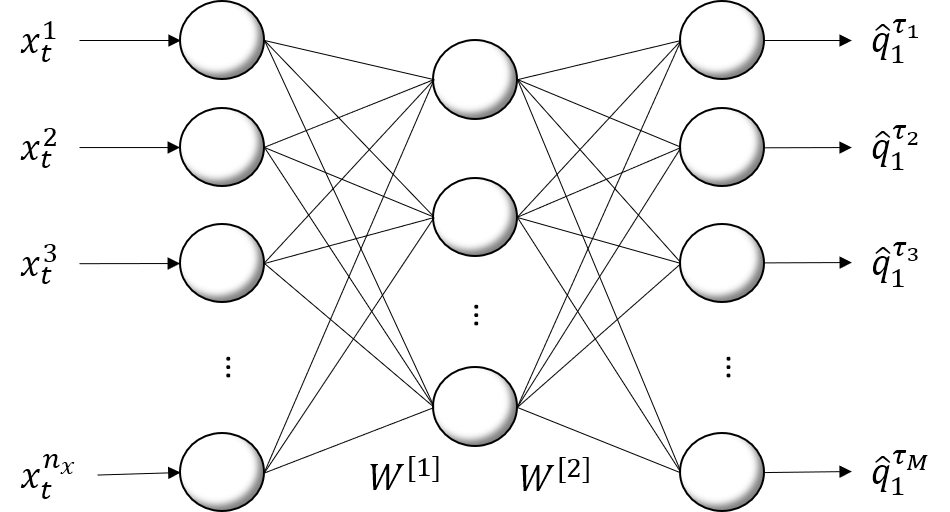}
	\caption{Schematic diagram of the smooth pinball neural network.}
	\label{f:net}
\end{figure}

\subsection{Smooth Quantile Regression}

\begin{figure}[t]
	\centering
	\includegraphics[width=.4 \textwidth]{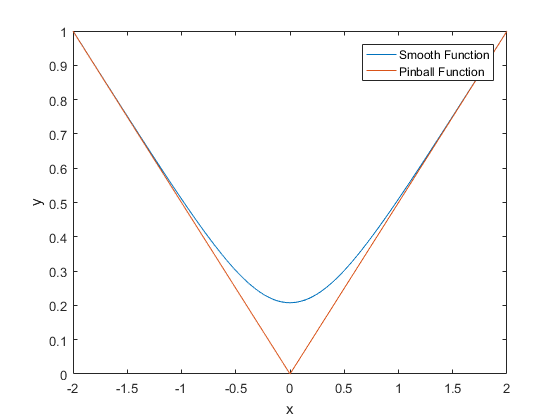}
	\caption{Pinball ball function versus the smooth pinball neural network with smoothing parameter $ \alpha =0.2 $.}
	\label{pinvsmooth}
\end{figure}

The smooth approximation \cite{zheng2011gradient} of the pinball function in Eq. \eqref{pinball} is given by
\begin{equation}
S_{\tau,\alpha}(u) = \tau u + \alpha \log \left( 1 + \exp \left(-\frac{u}{\alpha}\right) \right), 
\end{equation}

\noindent where  $ \alpha>0 $ is a smoothing parameter and $ \tau \in [0,1] $ is the quantile level we are  trying to estimate. In Fig. \ref{pinvsmooth} we see the pinball function with $ \tau = 0.5 $ as the red line and the a smooth approximation as the blue line with $ \alpha = 0.2 $. 
Zheng proves \cite{zheng2011gradient} that in the limit as $ \alpha \rightarrow 0^{+} $ that $ S_{\tau,\alpha}(u) = \rho_{\tau}(u) $. He also derives and discusses several other properties of the smooth pinball function. The smooth quantile regression optimization problem then becomes
\begin{equation} \label{Eq.smoothQR}
\min_{W,b} \frac{1}{N} \sum_{t=1}^{N} S_{\tau,\alpha} (y_t-\hat{q}_{t}^{(\tau)} ),
\end{equation}

\noindent where $ N $ is the number of training examples and $\hat{q}_{t}^{(\tau)} = W X_t + b$ where $W,b$ are the model parameters and $X_t$ is a vector of features at time $ t $. This form conveniently allows gradient based algorithms to be used for optimization.

\subsection{Smooth Pinball Neural Network}

For simplicity we describe here the construction of a single hidden layered SPNN for nonlinear multiple quantile regression, but SPNN can easily be extended to multiple hidden layers. In a single hidden layered SPNN the input layer consists of $ n_x $ number of input nodes and takes vector $ X_t $ of input features at time $ t $. The hidden layer consists of $ n_h $ number of hidden neurons and the output layer consists of $M$ number of output nodes corresponding to the estimated quantiles $ \hat{Q}_t = [\hat{q}_{t}^{(\tau_1)},...,\hat{q}_{t}^{(\tau_{M})}]^\top $ where $ \hat{q}_{t}^{(\tau_m)} $ is the $ \tau_m $ quantile level we want to estimate at time $t$. Every element in the first layer is connected to hidden neurons with the weight matrix $ W^{[1]} $ of size $ (n_x \times n_h) $ and bias vector $ b^{[1]} $of size $ (n_h \times 1) $. A similar connection structure is present in the second layer in the network between the hidden and output layers with $ W^{[2]} $ the output weight matrix of size $ (n_h \times M) $ and bias vector $ b^{[2]} $ of size $ (M \times 1) $. 

The input to hidden neurons is calculated, in vectorization notation, by $ Z_{t}^{[1]} =  W^{[1]} X_t + b^{[1]} $, the output of the hidden layer then uses the logistic activation function $ H_{t} = \tanh \left(  Z_{t}^{[1]} \right) $. The input to output neurons is then calculated by $ Z_{t}^{[2]} =  W^{[2]} H_t + b^{[2]} $, and the output layer uses the identity activation function $ \hat{Q}_t =  Z_{t}^{[2]} $.

The objective function for our SPNN model is then the smooth pinball approximation summed over $ M $ number of $ \tau $'s we are trying to estimate in the output layer. We also use L2 regularization on the network weights in the objective function to prevent over-fitting during training. The objective function for SPNN is then given by 
\begin{equation} \label{opti}
\begin{multlined}
E = \frac{\lambda_1}{2NM}\| W^{[1]} \|^{2}_{F}+ \frac{\lambda_2}{2NM}\| W^{[2]} \|^{2}_{F}+  \frac{1}{NM}\sum_{t=1}^{N} \sum_{m = 1}^{M} ...\\
\left[ \tau_m (y_t - \hat{q}_{t}^{(\tau_m)}) + \alpha \log \left( 1 + \exp \left( - \frac{y_t-\hat{q}_{t}^{(\tau_m)}}{\alpha} \right) \right) \right]. 
\end{multlined}
\end{equation}

\noindent where $ \| . \|_{F} $ is the Frobenius norm. Fig. \ref{f:net} shows a schematic diagram of our SPNN model with $ n_x $ number of input features and $ M $ number of quantile outputs.

Standard gradient descent with backpropagation can be used to train SPNN. Through this process we compute the gradient of the objective function $ E_t $ at each data point at time $ t $ with respect to $ W^{[1]}, b^{[1]}, W^{[2]} $ and $ b^{[2]} $. We start with the gradient with respect to the hidden-to-output weights $ W^{[2]}  $. In order to compute the gradient at time $ t $, we apply the chain rule in vector notation as follows
\begin{equation*}\label{wO}
\begin{multlined}
\frac{\partial E_t }{\partial W^{[2]}} = 
\frac{\lambda_2}{M} W^{[2]} + 
\frac{\partial E_t}{\partial \hat{Q}_{t}} \cdot \frac{\partial \hat{Q}_{t}}{\partial Z_{t}^{[2]}}
\cdot \frac{\partial Z_{t}^{[2]}}{\partial W^{[2]}}\\
=\frac{\lambda_2}{M} W^{[2]} +  \frac{1}{M} \left(\frac{1}{1+\exp \left( \frac{y_t-\hat{Q}_{t}}{\alpha} \right)} - T \right) H_{t},
\end{multlined}
\end{equation*}

\noindent where $ T = [\tau_1,...,\tau_m]^\top $ is a vector of all our $ \tau $'s. The gradient of $ b^{[2]} $ can be calculated similarly. Next we calculate the gradient of the objective function with respect to the weights of the first layer $ W_{[1]} $ as follows
\begin{equation*} \label{wI}
\begin{multlined}
\frac{\partial E_t}{\partial W_{[1]} } = \\ 
\frac{\lambda_1}{M} W^{[1]} + 
\left(  \frac{\partial E_t}{\partial \hat{Q}_{t}}  \cdot \frac{\partial \hat{Q}_{t}}{\partial Z_{t}^{[2]}}  \cdot \frac{\partial Z_{t}^{[2]}}{\partial H_{t} } \right) \cdot \frac{\partial H_{t}}{\partial Z_{t}^{[1]}} \cdot \frac{\partial Z_{t}^{[1]}}{\partial W^{[1]}}  \\
= \frac{\lambda_1}{M} W^{[1]} + \frac{1}{M}
\left( \frac{1}{1+\exp \left( \frac{y_t-\hat{Q}_{t}}{\alpha} \right)} - T \right) W^{[2]}\\
\left( 1-H_{t}^{2} \right) X_t
\end{multlined}.
\end{equation*}

\noindent The gradient of $ b^{[1]} $ can be calculated similarly. These gradients can then be directly used in many other gradient descent based optimization schemes. As such, we apply the Adam optimizer \cite{kingma2014adam}, an algorithm for first-order gradient-based optimization, to learn the parameters of SPNN. Adam has been shown \cite{kingma2014adam} to yield superior results compared to other gradient-based optimizers.

\begin{figure}[t]
	\centering
	\includegraphics[width=.35 \textwidth]{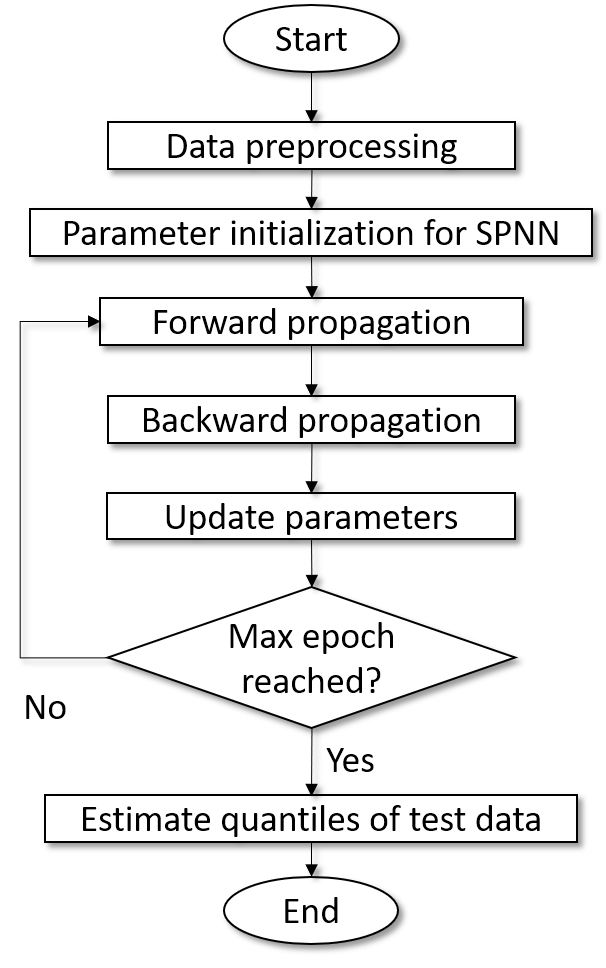}
	\caption{Flowchart of the steps taken when conducting a probabilistic forecast with SPNN.}
	\label{f:flowchart}
\end{figure}

\subsection{Noncrossing Quantiles}

In quantile regression normally a single quantile is estimated. To estimate multiple quantiles, one could be run QR to solve for different $ \tau $'s independently. However, in doing so, quantiles may cross each other which is not desirable since it violates the principle of monotonically increasing inverse density functions. To prevent this, we need to introduce constraints as per \cite{takeuchi2006nonparametric}. The condition $ 0< \tau_1<...<\tau_M $ are defined as the orders of $ M $ conditional quantiles to be estimated. To ensure these quantiles do not cross each other the following constraint is needed $ q_{t}^{(\tau_1)} \leq ... \leq q_{t}^{(\tau_M)}, \forall_t $. 

However, it is not easy to solve the neural network optimization problem with such constraints using gradient descent methods. One possible solution is proposed in \cite{cannon2018non} where a monotonic composite QRNN is presented that applies partial monotonicity constraints to the weights of the network and uses a stacked input matrix of covariates of size $N\times M$ with an added covariate $\tau_m$. This can add additional complexity to the network, by adding more parameters, so we propose a simpler alternative of applying a penalty term \cite{freund2004penalty} directly into the cost function. We define the non-crossing quantile penalty term $p$ as follows
\begin{equation}
p = c \sum_{t=1}^{N} \sum_{m=1}^{M} \left[ \max \left( 0, \epsilon -\left( \hat{q}_{t}^{(\tau_{m-1})}-\hat{q}_{t}^{(\tau_m)} \right) \right) \right]^2
\end{equation}
\noindent where $\hat{q}_{t}^{(\tau_0)} = 0 $, $\epsilon$ is the least amount that the two quantile should differ by, and $c$ is the penalty parameter with a high value. This penalty $p$ is added to the cost function in Eq. \ref{opti}. If the constraints are not violated no penalty is added to the cost function. If a lower quantile exceeds the value of a higher one, the squared difference of these two quantiles is added to the cost function as a penalty. A full model implementation flowchart is shown in Fig. \ref{f:flowchart}. First the data is preprocessed which includes deriving different input features, feature standardization, and partitioning the data into training and testing sets. Training of the model is conducted using gradient descent optimization method. After the max number of training epochs is reached the model is ready to be used on testing data for multiple quantile estimation. 

\section{Results and Discussions}\label{s:res}

To validate our model for probabilistic forecasting of wind power we utilize wind data from the publicly available Global Energy Forecasting Competition 2014 (GEFCom2014) \cite{hong2016probabilistic}. The goal of the wind component of GEFCom2014 was to design parametric or nonparametric forecasting methods that would allow conditional predictive densities of the wind power generation to be a function of input data which are numerical weather predictions (NWPs). Evaluation of predicted densities was done using the quantile score. Data is from the years of 2012 and 2013 from 10 wind farms titled Zone 1 to Zone 10. The predictors are NWPs in the form of wind speeds at an hourly resolution at two heights, 10m and 100m above ground level. These forecasts are for the zonal and meridional wind components (denoted U and V). It was up to the contestants to deduce exact wind speed, direction, and other wind features if necessary. These NWPs are from the exact locations of the wind farms. Additionally, power measurements at the various wind farms, with an hourly resolution, are also provided. All power measurements are normalized by the nominal capacity of their wind farm. The goal in forecasting is to learn to associate the provided NWPs (or derived features) with wind power. NWPs are provided for the forecasting horizon of one month, and it is up to a forecasting model to use those NWPs as input to predict quantiles at each future time step. 

\subsection{Benchmark Methods}

We use three standard \cite{sideratos2012probabilistic} and two advanced benchmark methods for density forecasting of wind power. The standard methods are the persistence model that corresponds to the normal distribution and is formed by the last 24 hours of observations, the climatology model that is based on all past wind power, and the uniform distribution that assumes all observations occur with equal probability. For our advanced benchmarks, we use a linear and nonlinear version of QR. The linear version is multiple quantile regression (QR) with L2 regularization, and nonlinear version is support vector quantile regression (SVQR) \cite{hwang2005simple} with a radial basis function kernel.

\subsection{Case Study Descriptions}

In the analysis of SPNN for forecasting wind power quantiles, we conduct studies with SPNN having one and two hidden layers denoted as SPNN1 and SPNN2. We study if the addition of a second hidden layer improves performance. Our SPNN model is a fully connected feedforward neural network, with rectified linear units for hidden activation functions, and it uses Adam for weight optimization \cite{kingma2014adam}. Default Adam parameters follow those provided in the original paper. The quality of the quantile estimates is sensitive to the hyperparameters of the network. SPNN has several hyperparameters that need to be chosen before training. Through empirical testing on training data, we found the following values as adequate for our model hyperparameters: 2000 training iterations, 200 batch size, 40 hidden nodes for SPNN1, 20 and 40 hidden nodes for SPNN2, 0.01 for the smoothing rate,  0.01 for each of the weight regularization terms, 1000 for the cross-over penalty term, and 0 for the cross-over margin.

For testing we conduct two case studies using the GEFCom2014 wind datasets. To ensure that our study is unbiased, we use for assessment the whole year of 2013. This dataset gives a total of $ 365\times24 = 8760 $ test samples for wind power forecasting per wind farm. The first case study uses wind data from Zone 1 and 2. We estimate quantile to produce prediction intervals with nominal coverage from 10\% to 90\% in increments of 10\%. The goal of this study is to evaluate the quantile and prediction interval estimates from SPNN in detail for reliability and sharpness. We also look at QVSS to see improvements between SPNN1 and SPNN2 use QR as the reference model. We also compare results to SVQR as it is the only other nonlinear quantile regression benchmark model. 

In the second case study, we estimate 99 quantiles on par with GEFCom2014. Results are derived for all ten wind farms in total, where we have 87,600 total test observations. For each test month, we are estimating 99 quantiles for 720 look ahead hours across ten farms. Results are derived across all Zones for QS, IS, ACE, and Sharpness. Given so much data we need a way to summarize results. Thus for every farm, we take the mean of all the evaluation scores across all Zones/months. In both case studies training is done using a sliding window of the previous twelve months to forecast the whole next month. Data from 2013 are used for hold out test sets. For instance, we start with predicting January 2013 using the past 12 months of 2012. After a month is predicted, the training window moves to incorporate new data and the prediction model is retrained to get a new prediction.

We run our case study on a computer with an Intel i7 6700 2.6 GHz, and 16 GB of RAM. For both studies, we use as input features the raw wind speed data at 10m and 100m for U and V directions. The only engineered features are four time features based on the hour of the day and day of the year: $ \cos( 2 \pi \frac{hour}{24}),\sin( 2 \pi \frac{hour}{24}),\cos( 2 \pi \frac{day}{365}),\sin( 2 \pi \frac{day}{365}) $. 

This is contrast with the winning teams from GEFCom2014 who each used dozens of engineered features including lagged data, data from neighboring wind farms, and more complex features such as derived wind speeds, wind direction, wind energy, wind shear, direction differences between 10m and 100m, etc. Most of the winning teams in GEFCom2014 conducted heavy manual feature engineering to reduce the quantile score throughout the competition. The goal of our study is not custom feature engineering, which might result in better scores, but to highlight the effectiveness of SPNN in creating its own latent features via its hidden layers, and to showcase the feasibility of our method as a robust probabilistic forecasting model.

\subsection{Case Study 1}

\begin{figure}[t]
	\centering
	\includegraphics[width=0.5 \textwidth]{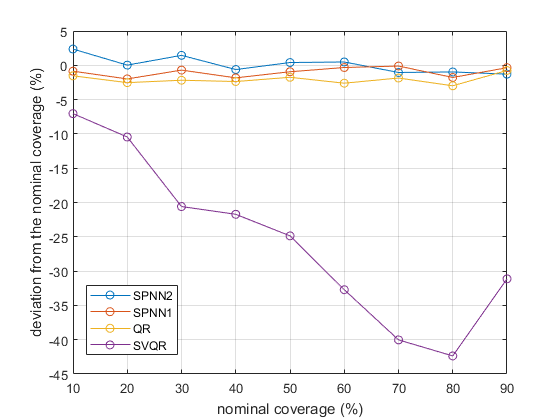}
	\caption{Reliability of prediction intervals from Zone 1 measured by the frequency of observation falling with each interval.}
	\label{f:reliability_zone1}
\end{figure}

\begin{figure}[h]
	\centering
	\includegraphics[width=0.5 \textwidth]{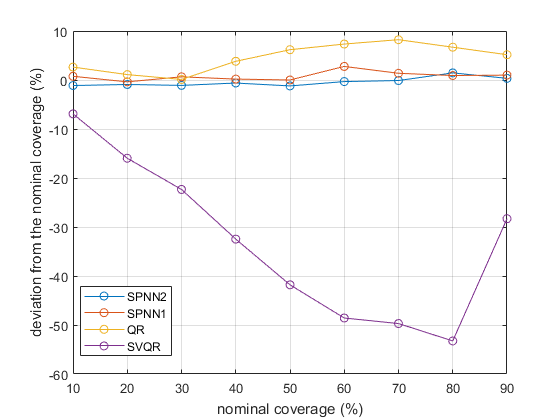}
	\caption{Reliability of prediction intervals from Zone 2 measured by the frequency of observation falling with each interval.}
	\label{f:reliability_zone2}
\end{figure}

For this first case study, quantiles are computed to form predictive intervals. Each prediction interval is estimated to have a future observation of wind power within a lower and upper bound for a given probability or nominal coverage rate. As previously mentioned, we estimate quantile to produce prediction intervals with nominal coverage from 10\% to 90\% in increments of 10\%. We estimate intervals for SPNN1, SPNN2, QR, and SVQR. The difference between the nominal coverage rates and the observations for Zone 1 are shown in Fig. \ref{f:reliability_zone1}. This reliability diagram showcases results similar to the ACE score. It can be see that SPNN2 has the lowest deviation from the nominal coverage with SPNN1 and QR coming second and third with result magnitudes ranging from -3\% to -0.3\%. SVQR has a very poor coverage with deviations as high as -40\%. This can be attributed to having too tight intervals and over-fitting. In Fig. \ref{f:reliability_zone2} we showcase reliability results from Zone 2. Similarly to Zone 1, SPNN2 yields intervals with a deviation close to 0, while SVQR continued to have a poor coverage.

\begin{figure}[h]
	\centering
	\includegraphics[width=0.5 \textwidth]{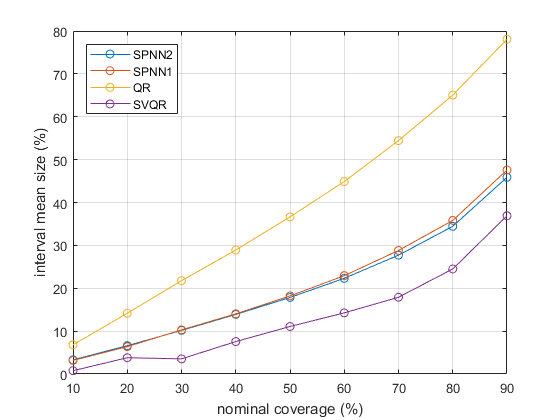}
	\caption{Sharpness of prediction intervals for Zone 1 measured by the interval mean size.}
	\label{f:sharpness_zone1}
\end{figure}

\begin{figure}[h]
	\centering
	\includegraphics[width=0.5 \textwidth]{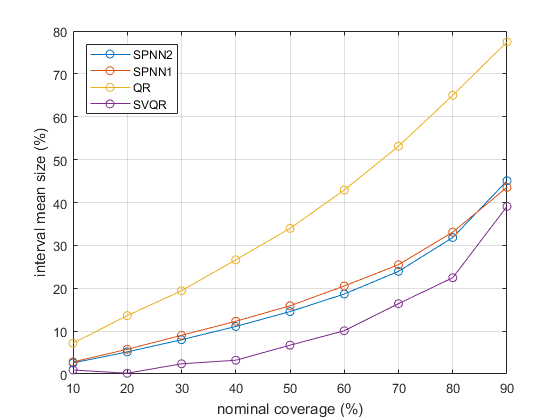}
	\caption{Sharpness of prediction intervals for Zone 2 measured by the interval mean size.}
	\label{f:sharpness_zone2}
\end{figure}

Sharpness is the other important statistic that we look at for individual predictive intervals which is calculated independent of observations. Measured as the mean interval size as described in Eq. \ref{e:sharp}, it demonstrates the usefulness of predictions. Ideally, we would like to have intervals as small as possible but too small and observations may fall outside the intervals. Thus, too wide and too narrow intervals providing poor forecasts. Sharpness needs to be analyzed together with reliability to ensure robust predictions. In Fig. \ref{f:sharpness_zone1}, we see the mean interval sizes for each coverage level for Zone 1. QR resulted in having the widest intervals and SVQR having the narrowest intervals. With such narrow intervals SVQR was not able to capture the observations which indicated in its reliability diagram. In Fig. \ref{f:sharpness_zone2} we see similar results for Zone 2. Our proposed method, SPNN1 and SPNN2, were able to estimate effective sized intervals that resulted in high reliability with good sharpness.

\begin{figure}[h]
	\centering
	\includegraphics[width=0.5 \textwidth]{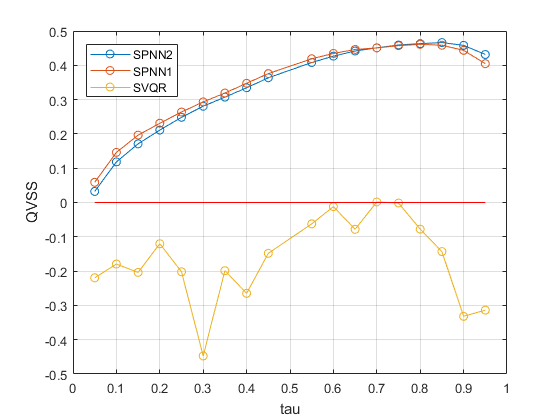}
	\caption{QVSS measured relative performance of SPNN2, SPNN1, and SVQR to QR on Zone 1 dataset.}
	\label{f:qvss_zone1}
\end{figure}

\begin{figure}[h]
	\centering
	\includegraphics[width=0.5 \textwidth]{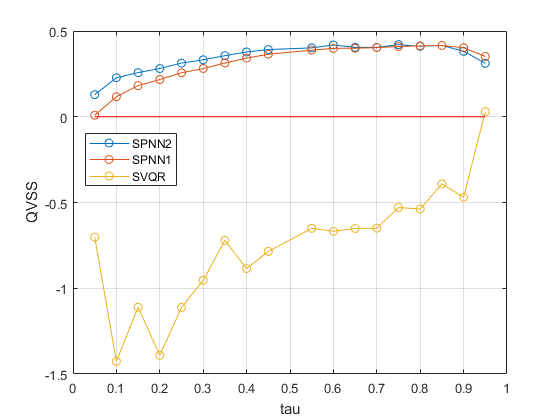}
	\caption{QVSS measured relative performance of SPNN2, SPNN1, and SVQR to QR on Zone 1 dataset.}
	\label{f:qvss_zone2}
\end{figure}

As a last evaluation, we look at the performance of the individual quantiles that formed the prediction intervals of this case study. We do this using QVSS to analyze relative performance gain relative to a reference benchmark model. Here we use quantile regression for the reference model and we study if the nonlinear quantile regression models, SPNN and SVQR provide any improvements over QR. In Fig. \ref{f:qvss_zone1}, we report the QVSS across the 18 quantiles for the three nonlinear methods. SPNN1 and SPNN2 provide a clear performance increase with respect to QR. For quantiles with a nominal probability less then 0.7, we see SPNN1 having a small lead over SPNN2. While SPNN2 shows a small lead for quantiles with $ \tau > 0.7 $. Not surprisingly, SVQR shows a decreased negative performance over QR, indicating its inability to extract meaningful features from the raw data for Zones 1 and 2. In Fig. \ref{f:qvss_zone2}, we see similar QVSS results for Zone 2, but with SPNN2 showing a small lead over SPNN1 for quantiles with $ \tau < 0.7 $.

\subsection{Case Study 2}

In our second case study we analyze a higher number of estimated quantile (99) across all wind farms for all 12 test months to ensure an unbiased assessment of SPNN relative to the benchmark models. Due to the large number of quantiles and wind farms, instead of forming reliability or sharpness diagrams for individual PIs and QVSS diagrams for individual quantiles, we instead look at box plots and report the distribution of evaluation results including QS, IS, ACE, and Sharpness.

\begin{figure}[t]
	\centering
	\includegraphics[width=0.5 \textwidth]{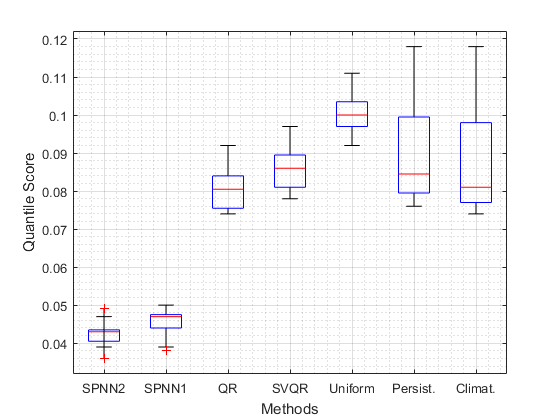}
	\caption{Box plot of quantile score evaluation across all datasets.}
	\label{f:barQS}
\end{figure}

\begin{figure}[t]
	\centering
	\includegraphics[width=0.5 \textwidth]{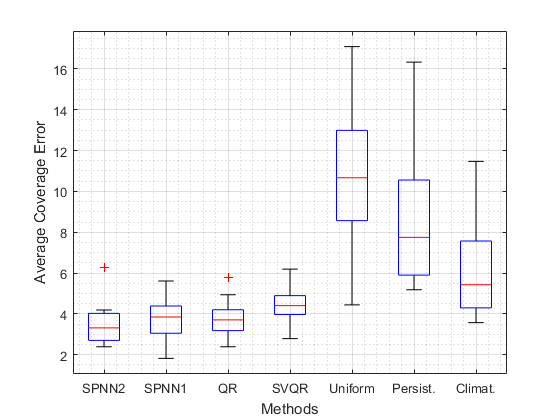}
	\caption{Box plot of average coverage error evaluation across all datasets.}
	\label{f:barACE}
\end{figure}

\begin{figure}[t]
	\centering
	\includegraphics[width=0.5 \textwidth]{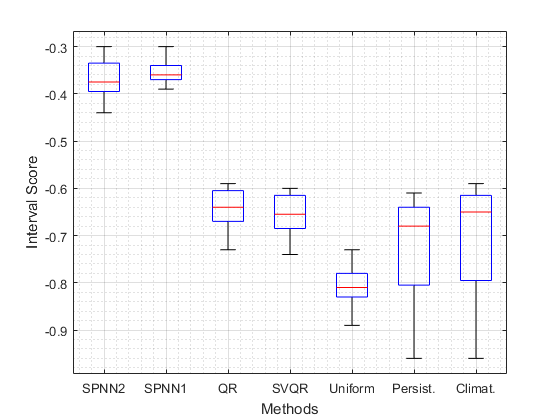}
	\caption{Box plot of interval score evaluation across all datasets.}
	\label{f:barIS}
\end{figure}

\begin{figure}[t]
	\centering
	\includegraphics[width=0.5 \textwidth]{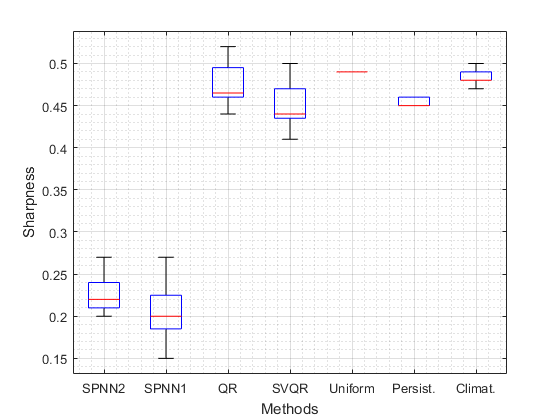}
	\caption{Box plot of sharpness evaluation across all datasets.}
	\label{f:barS}
\end{figure}

In Fig. \ref{f:barQS} we report the QS metric for SPNN and the five benchmark methods. We see that SPNN2 had the lowest QS range from 0.036 to 0.047 with SPNN1 being a close second. The other benchmarks had a QS in the range of 0.075 to 0.011. Inspecting the coverage analysis of our prediction intervals with the ACE score in Fig. \ref{f:barACE}, we see that SPNN overall has the lowest ACE with SPNN2 having a median value lower then SPNN1. The uniform benchmark produced a wide range for the ACE score due to having fixed size intervals across all zones and months, while SPNN2 had the narrowest range of ACE scores. Looking at the sharpness of PIs with the interval score in Fig. \ref{f:barIS} and general sharpness score in Fig. \ref{f:barS}, we see that SPNN has the sharpest intervals across all farms. The persistence and climatology methods yielded a wide distribution for the interval score but narrow one for sharpness. SVQR in contrast to the first case study did not calculate narrow intervals when estimating 99 quantiles.

Since both QS and IS also measure skill, we can say that SPNN was able to produce the highest quality estimates from all methods. An interesting observation is the SPNN is designed to produce optimal quantile estimates and that indirectly it also produces adequate interval forecasts. If the primary goal is to reduce ACE and IS as best as possible, alternative loss functions that incorporate prediction interval coverage and width functions can be used. However, while not directly optimizing for coverage or sharpness, SPNN does produce superior results from the advanced benchmarks multiple quantile regression and support vector quantile regression. 

Lastly, we compare the mean QS of our proposed method to the final quantile scores for the top teams in the GEFCom2014 as originally reported in \cite{hong2016probabilistic}. We note again that the top teams used a wide range of engineered features while we used raw wind speed data along with time as input to our model. The winning team in GEFCom2014 was kPower with a mean QS of 0.038. Our method SPNN2 has a close mean QS of 0.042 which would qualify SPNN to be in the top winning teams. Comparing the results from the four box plots, we see the robust prediction ability of the proposed SPNN prediction method. Additionally, for all the runs across months and farms, the preassigned PI coverage levels are satisfied which implies that the constructed PIs cover the target values with a high probability and with the lowest QS and IS.

\begin{figure}[t]
	\centering
	\includegraphics[width=0.45 \textwidth]{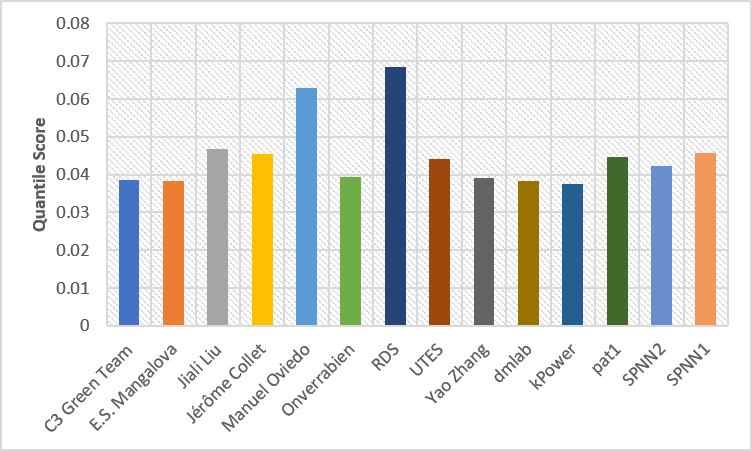}
	\caption{Bar plot of SPNN2 and SPNN1 mean quantile score across all wind data compared to the performance of the top teams in GEFCom2014 Wind Track.}
	\label{f:bar}
\end{figure}

\section{Conclusion} \label{s:con}

Wind power forecasting is crucial for many decision-making problems in power systems operations and is a vital component in integrating more wind into the power grid. Due to the chaotic nature of the wind, it is often difficult to forecast. Uncertainty analysis in the form of probabilistic wind prediction can provide a better picture of future wind coverage. This paper proposes a novel approach we call SPNN for probabilistic wind forecasting using a neural network with a smooth approximation to the pinball ball loss function in estimating multiple quantiles. 

We also introduce non-crossing constraints in the form of a smooth penalty in the loss function. This is done to ensure multiple quantiles can be estimated simultaneously without overlapping each other. We verify the effectiveness of our SPNN model  with the dataset of the Global Energy Forecasting Competition 2014. We compare forecasts to standard and advanced benchmarks and employ standard quantile score, reliability, and sharpness metrics. Our results show superior performance across the prediction horizons, which verify the effectiveness of the model for forecasting while preventing estimated quantiles from overlapping. 

Our SPNN method has the potential to be applied to a variety of domains for probabilistic forecasting or multiple quantile estimations. Future work will look into applying SPNN to forecast solar and ocean wave power, to test its effectiveness across different renewable energies, and on electricity pricing and load demand for smart grid applications. In this study, we trained our model using NWP data. Another problem to study is very short-term probabilistic forecasting using only past wind power data. Future work can also then look into expanding the SPNN model for providing full predictive densities given lagged past data of power only. 

\bibliographystyle{IEEEtran}
\bibliography{mybib}

\end{document}